# Structure and magnetism in ultrathin iron oxides characterized by low energy electron microscopy

B Santos[1,2], E Loginova[3], A Mascaraque[4], A K Schmid[5], K F McCarty[3] and J de la Figuera[1,2]

[1] Universidad Autónoma de Madrid, Madrid 28049, Spain

[2] Instituto de Química-Física "Rocasolano", CSIC, Madrid 28006, Spain

[3] Sandia National Laboratories, Livermore, California 94550, USA

[4] Universidad Complutense de Madrid, Madrid 28040, Spain

[5] Lawrence Berkeley National Laboratory, Berkeley 94720, USA

PACS: 68.37.Nq, 68.47.Gh, 75.70.-i

Email: juan.delafiguera@iqfr.csic.es

## Abstract

We have grown epitaxial films a few atomic layers thick of iron oxides on ruthenium. We characterize the growth by low energy electron microscopy. Using selected area diffraction and intensity vs. voltage spectroscopy, we detect two distinct phases which are assigned to wüstite and magnetite. Spin polarized low energy electron microscopy shows magnetic domain patterns in the magnetite phase at room temperature.

## 1 Introduction

The rich history of scientific interest in iron oxides has been fuelled by valuable applications taking advantage of catalytic, electronic, and magnetic properties of these materials[1]. Exploration of the diverse structural and electronic properties of iron oxide surfaces, thin films and nanostructures is a fruitful research area[2]. In particular, the possibilities of half-metal electronic structure[3, 4, 5] and multiferroic properties[6, 7, 8] of magnetite have been attracting interest.

Current research in this field benefits from recent advances in experimental technique, including the development of methods to grow highly perfect, epitaxial Fe oxide films on different substrates in ultra-high-vacuum (UHV) conditions (see [9, 10] and reference therein). The possibility to fabricate oxides through hetero-epitaxy is helpful in several ways, first because a suite of standard surface science techniques is available for in-situ characterization, and second, because epitaxy opens the door to tailoring materials properties through epitaxial strain[11], nano-scale self-assembly, etc.

The most common iron oxides phases, in order of increasing Fe oxidation state, are FeO (wüstite), $Fe_3O_4$ (magnetite), $\gamma- Fe_2O_3$ (maghemite) and $\alpha- Fe_2O_3$ (hematite). The stability ranges of these oxides as a function of temperature and oxygen pressures has been computed and summarized in phase diagrams[12]. When growing the materials in thin-film form by annealing in pressures of molecular oxygen (below $10^{-6}$ torr), the phases most often reported are wüstite (FeO) and magnetite ($Fe_3O_4$).

Wüstite is an antiferromagnet with a Néel temperature of around 200 K. Below the Néel temperature a magnetocrystalline distortion results in a rhombohedral structure[1]. Magnetite, the oldest known magnetic material, is ferrimagnetic with an extremely high Curie temperature of 850 K. Below 120 K it undergoes a first order phase transition (Verwey transition) at which the conductivity decreases by two orders of magnitude. Also, the lower temperature phase is ferroelectric[6, 7]. The precise origin of this transition is a subject of active research and discussion[13, 14].

Iron oxides have been grown in a thin film form on several metal surfaces, including Pt(111)[10], Cu(100)[15], Ru(0001)[16], Au(111) and Ag(111)[17]. The growth is typically performed by depositing a few atomic layers of iron and oxidizing afterwards. When using molecular oxygen in UHV as oxidizing agent, the resulting oxides tend to follow a common sequence: the thinnest films are interpreted as a form of FeO(111), and these films wet the metal substrates. The maximum FeO thickness that can be grown in this way depends on the substrate, with up to four FeO layers reported on Ru(0001)[16]. When a larger amount of material is deposited, a new phase begins to nucleate in the form of 3-dimensional (3D) crystallites. Density and average size of the 3D crystallites depends on details of the experimental conditions. Continuous films of this phase can also be grown, and the excellent fit between computed spectra and low energy electron diffraction (LEED) intensity-vs-energy (IV) data suggests that these films are composed of magnetite[10]. Mössbauer spectroscopy further confirms this interpretation[18]. Beyond the oxidation state of magnetite, the growth of maghemite films can be achieved[19] by using atomic oxygen as a stronger oxidation agent than $O_2$.

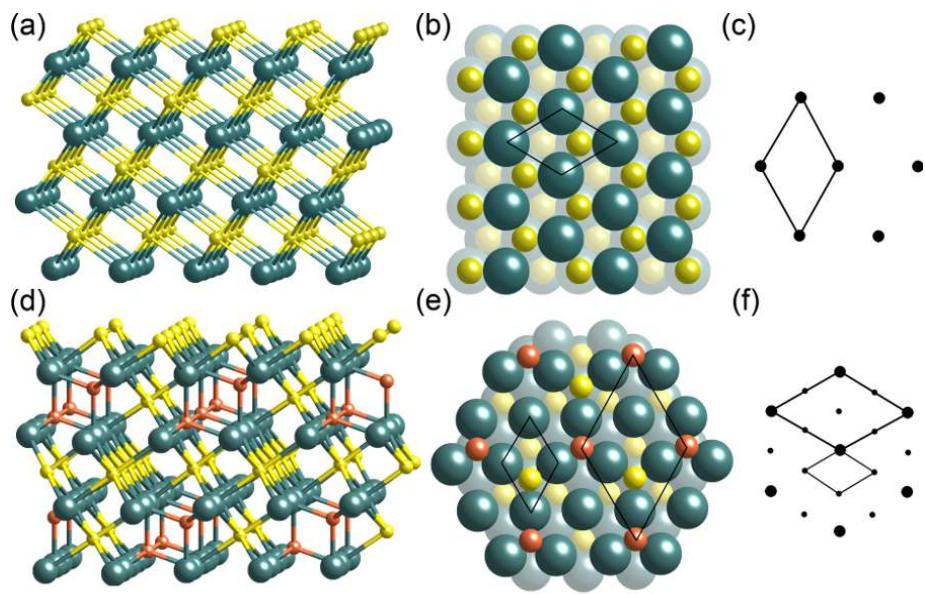

Figure 1: Schematics of wüstite (a-c) and magnetite (d-f). For each phase, we show a side view, the top (111) view and the LEED pattern expected for unreconstructed surfaces. In wüstite, compact Fe (yellow atoms) and O (gray atoms) hexagonal planes alternate, with the same unit cell in each (marked on the top view and reciprocal space schematics). In magnetite, oxygen (gray) atoms are arranged in an fcc structure, while Fe planes (yellow and red balls indicate, respectively, the octahedrally and tetrahedrally coordinated atoms) are arranged in a more complicated sequence, with a (2x2) unit cell relative to oxygen hexagonal planes. Both the O and Fe unit cells are show on the top view and the reciprocal space schematic.

For this paper, it is useful to summarize key aspects of the structures of iron oxides. An instructive view has been described by Cornell and Schwertmann in ref. [1]. The oxygen anions have a larger size that the iron cations, and the structure of most oxides can thus be pictured by assuming that the oxygen anions form a locally distorted compact structure, either fcc or hcp. The smaller iron cations occupy either tetrahedrally (A) or octahedrally (B) coordinated sites between the oxygen atoms. The cations in these oxides are either $Fe^{2+}$ and/or $Fe^{3+}$ (The larger $Fe^{2+}$ cations occupy the larger octahedral sites.) The oxygen distances are not too different among the different oxide structures and correspond to a nearest-neighbour distance of $\approx 3.0$ Å. (3.04 Å, 2.9 Å, and 3.3 Å for $Fe^{2+}$ in octahedral coordination, $Fe^{3+}$ in octahedral coordination, and $Fe^{3+}$ in tetrahedral coordination, respectively).

FeO crystallizes in the NaCl structure, i.e., it is composed along the [111] direction of alternating Fe and O planes with the $Fe^{2+}$ cations occupying all the available octahedral sites between oxygen atoms (see figure 1a). In the bulk the phase is non-stoichiometric, with some $Fe^{2+}$ atoms in the $Fe^{3+}$ oxidation state, and a corresponding change in lattice spacing. The FeO phase has a lattice parameter of 4.28-4.31 Å which translates into an in-plane unit-cell for both the Fe and O planes of 3.02-3.05 Å in the (111) layers.

Magnetite crystallizes in the inverse spinel structure, and presents a mixture of $Fe^{2+}$ and $Fe^{3+}$ ions in interstitial sites within an fcc-like anion lattice (see figure 1d-f). While tetrahedral (A) sites are populated by $Fe^{3+}$ ions, the octahedrally coordinated sites (B) are occupied by equal numbers of $Fe^{2+}$ and $Fe^{3+}$ cations. Along the [111] direction, hexagonal layers of oxygen with an in-plane lattice spacing close to 2.97 Å are alternatively separated by either a layer of $Fe^{2+}$ and $Fe^{3+}$ in octahedral coordination together with 1/4 of vacancies, or by three layers with only 1/4 occupancy each (corresponding to, respectively, tetrahedral, octahedral and tetrahedral sites). Maghemite, $\gamma-Fe_2O_3$, presents the same structure as magnetite but with the $Fe^{2+}$ ions replaced with vacancies[1].

In this work we present a low-energy electron microscopy (LEEM[20]) characterization of the first stages of the growth of iron oxides on ruthenium. We exploit the advantages of LEEM to observe in real time the deposition of iron under molecular oxygen at high substrate temperatures. We characterize the oxides by low-energy electron diffraction and low-energy electron reflectivity. Finally we present spin-polarized LEEM images that show the magnetization states of iron oxide islands and films.

## 2 Experimental details

We performed the growth and characterization of our samples in-situ in two different ultrahigh vacuum low-energy electron microscopes. Both instruments have facilities for in-situ heating (up to 2300 K) and cooling (down to 100 K) the samples while recording images at up to video rate. One is a conventional LEEM[20], which can acquire selected-area diffraction[21, 22] data from regions as small as 0.5 μ m in diameter. The second instrument is equipped with a spin-polarized electron gun (SPLEEM[23]), providing the possibility to image magnetic domain structures[24, 25]. The instrument has a spin-manipulator to change the electron-beam polarization direction to any desired orientation[26]. By forming pixel-by-pixel difference images from pairs of images acquired with opposite spin polarization, non-magnetic contrast effects are suppressed while contrast due to magnetization of the sample is enhanced. It is possible to acquire sets of three such difference-images using orthogonal quantization axes (usually the direction perpendicular to the surface plus two orthogonal in-plane directions), and from comparison of magnetic contrast in the sets of images the 3D components of the magnetization vector in the sample surface can be mapped

[27].

All our films were grown on Ru(0001) single crystals. The procedure we used for cleaning the substrates is described elsewhere[25]. Imaging in real time the growth of Fe on Ru at 520 K substrate temperature allowed us to calibrate the Fe flux from the e-beam evaporators by measuring the time needed to complete the first monolayer. The rates we used were typically ≈ 0.3 ML/min, and total dose of Fe used in individual film depositions was estimated from carefully controlled deposition times.

## 3 Results and discussion

To follow the first stages of the iron oxide growth, we imaged the Ru(0001) surface while depositing iron at a pressure of molecular oxygen of typically $10^{-6}$ torr. We first show the growth of iron at low and high temperatures, and then the results of the in-situ observation of oxide growth. We will then present the LEED characterization, followed by the reflectivity measurements, and end with the spin-polarized LEEM experiments.

*3.1 Real-time LEEM: iron growth on ruthenium*

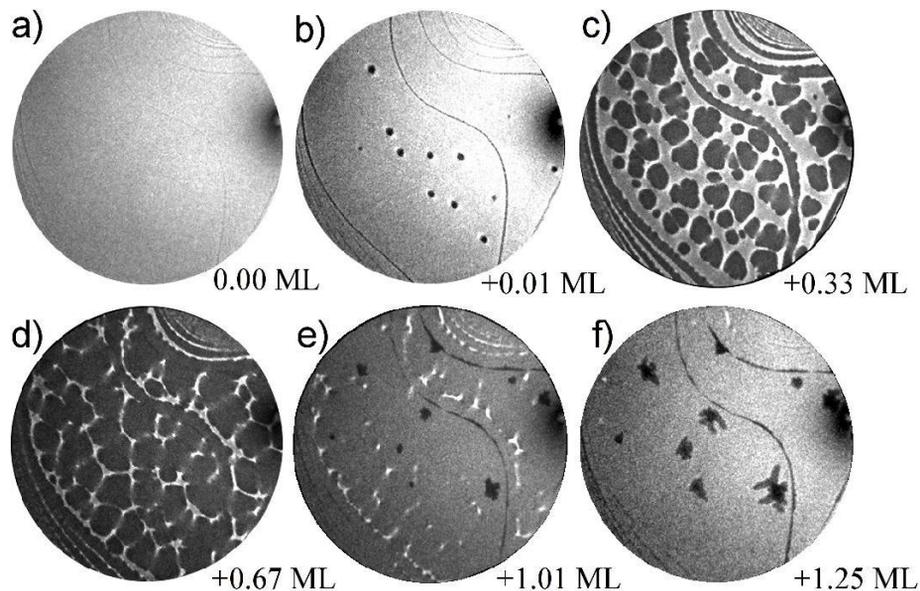

Figure 2: Growth of Fe on Ru(0001) in vacuum at 521 K. a-f) LEEM images extracted from a sequence acquired in real time during iron growth. The smooth dark lines in a) and b) mark atomic steps in the substrate. Fe islands 2 ML or thicker

appear dark in f). The field of view (FOV) is 10 μm, and the electron energy is 4 eV. Labels give the total amount of deposited Fe in ML.

The growth of iron on ruthenium proceeds in a similar way to other transition metals deposited onto refractory metal substrates. An example is shown in figure 2. The first layer grows wetting the substrate, nucleating into rounded islands (figure 2b-d). The monolayer Fe film produces a LEED pattern (not shown) with the spots at the same position as the Ru substrate, indicating that Fe grows with the same in-plane lattice spacing. Subsequent layers grow by forming 3-dimensional islands. Close inspection of the image reproduced in figure 2f already reveals the onset of 3D island growth, the dark islands with ramified shapes already contain smaller and darker regions with higher thickness. The Fe/Ru system has been studied previously using several techniques[27,28], and it has been reported that interdiffusion between the substrate and film is kinetically suppressed at temperatures below 600 K.

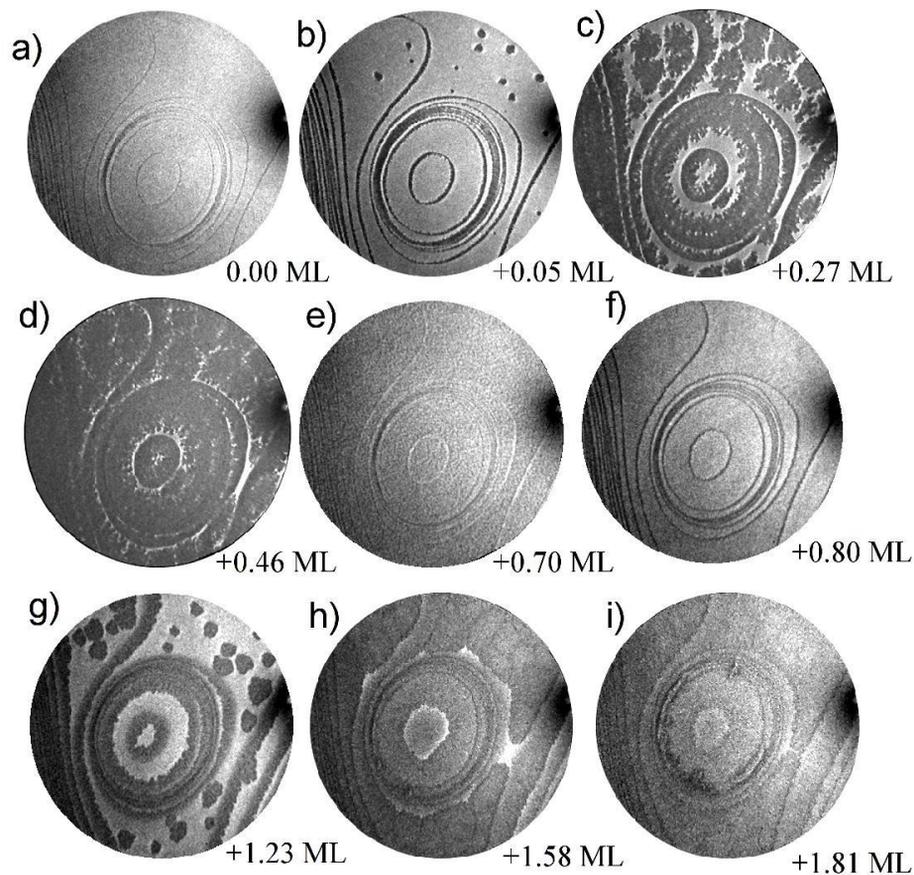

Figure 3: Growth of Fe on Ru(0001) at 843 K. At this temperature, alloying of the Fe and Ru causes a different growth mode than at lower temperature, as shown in figure

2. a-f) LEEM images extracted from a sequence acquired in real time during iron growth. The field of view (FOV) is 10 μm, and the electron energy is 4 eV. Labels give the amount of deposited Fe in ML.

Above approximately 800 K, interdiffusion between Fe and the substrate is activated. At high temperature, about two Fe monolayers[28] remain at the surface, with additional iron alloying into the bulk of the ruthenium substrate. One of our high-temperature (843 K) Fe/Ru growth experiments is shown in figure 3. The first-layer islands are highly dendritic (figure 3c) in contrast with 1 ML islands grown at lower temperature, which have compact shapes (figure 2c). This observation is consistent with thermally activated alloying: in heteroepitaxial systems that do not form surface alloys, one often observes that island shapes become increasingly more compact as a function of higher temperature (this is interpreted as step-edge smoothing resulting from thermally activated step-edge or surface diffusion). High-temperature dendritic growth observed in other metal-on-metal systems has been attributed to result from surface alloying[29,30]. In the case of Pd/Ru[30], it was shown that the formation of a dilute surface alloy impedes island growth and gives rise to instability of step-flow growth at island edges. This instability then causes the growth of extremely non-compact island shapes. We speculate that a similar process is at work in the Fe/Ru system. Continuing Fe deposition beyond one monolayer thickness, we observe that the second layer grows flat, in contrast with lower temperature growth (compare figure 2f and figure 3g). During growth of the second monolayer, we also observe a noticeable change in electron reflectivity (image grey level) as a function of distance to substrate steps. This heterogeneous electron reflectivity is reminiscent of what is observed in Pd/Cu(100)[29,31], where a detailed reflectivity study indicated that the heterogeneity is due to a step-overgrowth alloying mechanism.

Consistent with previous reports[27] we interpret our observations of Fe/Ru growth at 843 K to indicate that a certain extent of interdiffusion between Fe and the Ru substrate is activated in this temperature regime.

*3.2 Real-time LEEM: iron-oxides grown on ruthenium*

Alloying is a concern for iron oxide growth on ruthenium because, in the presence of alloying, interface sharpness is likely to be affected and, possibly, iron oxides grown at high temperature may contain a certain amount of Ru. However, as will be discussed in this section, we find that simple iron-oxide film growth recipes tend to result in the simultaneous presence of several oxide phases. In order to be able to apply micro-diffraction techniques for the identification of these phases, it is helpful when individual precipitates are not too small. In our iron oxide growth experiments, we found that achieving sufficiently large precipitate sizes to allow unambiguous results from our micro-diffraction techniques required treatments in a temperature range where alloying is no longer kinetically suppressed. While this implies that we cannot exclude the possibility that our iron oxide films contain some ruthenium, we tend to believe that the extent of such interdiffusion is limited. First, it is known that at high temperature a segregated bilayer of iron on ruthenium is stable[27,28]. Secondly, ruthenium is a noble metal while iron forms very strong bonds with oxygen. Therefore, one might guess that the presence of oxygen might lessen the tendency for alloying.

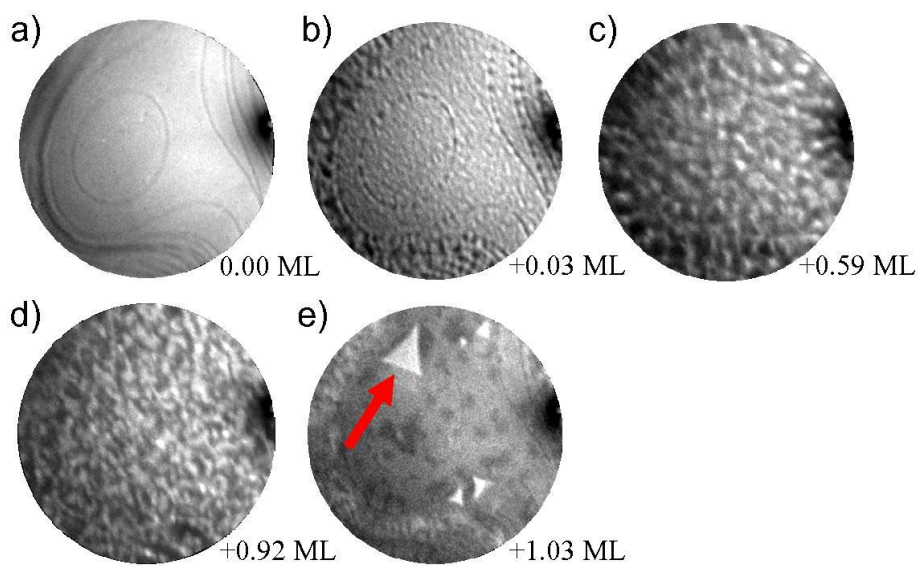

Figure 4: Growth of Fe in $10^{-6}$ torr oxygen at 893 K. a-e LEEM images extracted from a sequence acquired in real time during iron growth. The field of view (FOV) is 10 μm. A triangular island is marked with a red arrow in e. Labels give the amount of deposited Fe in ML. The electron energy is 19.5 eV.

To minimize alloying of Fe and Ru, common procedures are to either dose Fe at low temperature and then anneal in oxygen, or directly dose Fe in an oxygen atmosphere. Dosing in oxygen allows us to monitor the oxide growth by LEEM imaging. This is interesting because, in contrast to many detailed studies of nucleation and growth of metal and semiconductor films, there is a distinct shortage of in-situ, real time imaging studies of oxide growth. We present in figure 4a-e several frames extracted from a sequence taken while depositing iron on a Ru(0001) surface under $10^{-6}$ torr oxygen pressure. The substrate temperature was 893 K. From the initial stages (figure 4a-d), the growth proceeds by very small islands, not clearly resolved in the images. The islands seem to coarsen to the point where, close to a total dose corresponding to 1 ML of iron, there is a sudden change and the surface becomes more uniform. At about the same time (see figure 4e) triangular islands start to grow. Density of the triangular islands is low, and on the surface regions between the islands, irregular patches with slightly different electron reflectivity appear. This morphology remains stable when the sample is cooled to room temperature (RT) and the oxygen atmosphere is removed (see figure 5a). The two types of areas of the surface, triangular islands versus the regions between the islands, are clearly different: small-area LEED shows that the diffraction patterns are different, as reproduced in figure 5c,d. The morphology of this film appears to remain stable during brief post-annealing in oxygen (not shown) or in vacuum up to 1206 K (see figure 5b). The only annealing-induced change we observe is that the regions between the islands become more uniform. In the following sub-section, we argue that the triangular islands and the surrounding film are different phases of iron oxide.

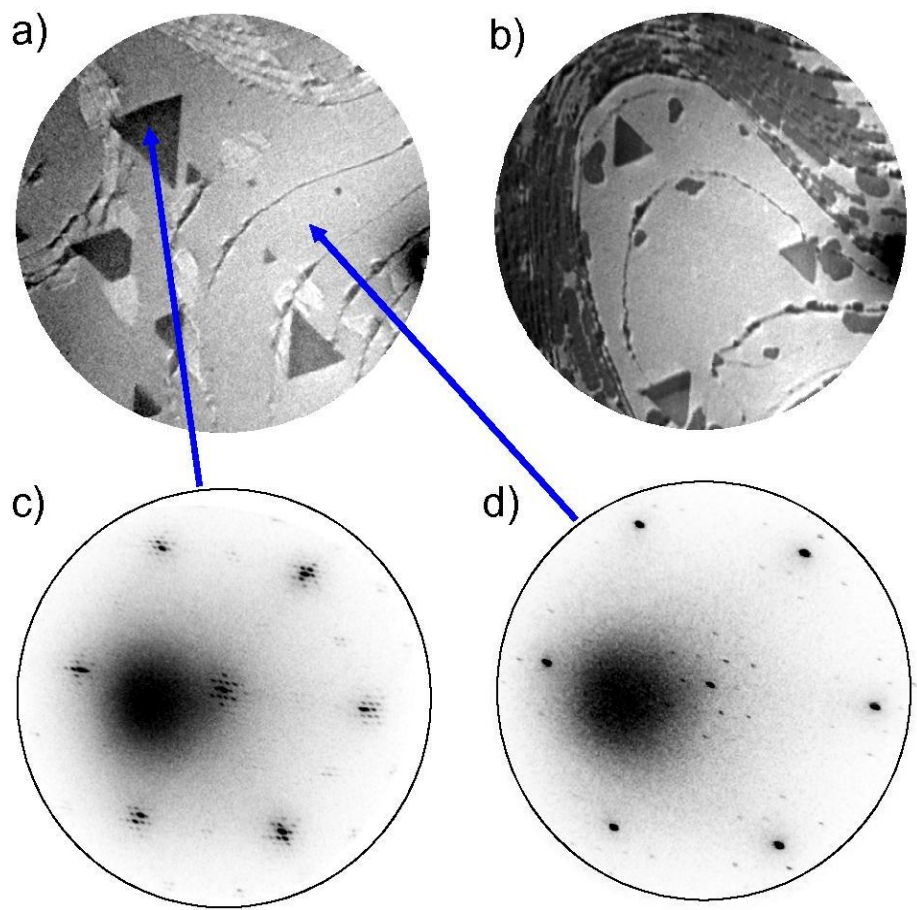

Figure 5: a,b) LEEM images of the iron oxide film described in figure 4, after annealing to 1206 K for a few minutes. The reversed contrast between triangular islands and the surrounding areas in these images when compared with figure 4e is due to differences in the electron energy used [27 and 36 eV respectively for images a) and b)]. c) LEED pattern (45 eV) from the marked triangular island. d) LEED pattern (45 eV) from the film areas between the islands.

Two characterization techniques available within LEEM instruments are low energy electron diffraction and low energy reflectivity. The use of an illumination aperture in the beam trajectory allows us to acquire diffraction patterns from areas of the surface as small as a fraction of a micron. In this way we can separate the diffraction patterns of the different surface areas observed in LEEM images, as shown in figure 5.

*3.3 LEED and reflectivity characterization*

We present in figure 6 a detailed view of the patterns observed on the iron oxide films. On the clean Ru substrate (see figure 6b), the diffraction pattern from a single terrace shows three-fold symmetry, as expected for an hcp substrate[32]. In oxide films that were prepared similarly to the ones shown in figure 4 and figure 5, most of the substrate is covered with a continuous layer which has the type of LEED pattern reproduced in figure 6c-d. This diffraction pattern has three-fold symmetry, albeit with a substantially larger unit cell compared to the Ru substrate, as indicated by the smaller separations in reciprocal space of the most intense spots shown in figure 6d. The pattern is as sharp as that of the substrate, indicating very good order in the oxide film. Quantitative comparison of spot positions shows that the unit cell of the oxide film is aligned with the substrate, and has a lattice spacing of $d=3.02\pm0.2$ Å. The margin of error is estimated by averaging over the equivalent integer beam positions over several experiments (the error estimate is dominated by distortions in the electron optics). In addition to the main diffracted beams, there are satellite spots around both the specular and the integer beams. These satellite spots arise from a superstructure produced by the coincidence pattern of the iron oxide film and the ruthenium substrate. The satellite spots correspond to a real-space periodicity of $18.5\pm2$ Å. In a previous study based on LEED and STM experiments, Ketteler et al.[16] reported unit cell and superstructure periodicities of 3.08 Å and 21.6 Å, respectively, and interpreted the structure as a single layer FeO (wüstite) phase. Noting the good agreement of our LEED results with their measurements, we suggest that on our samples the film regions between the triangular islands are also FeO wüstite.

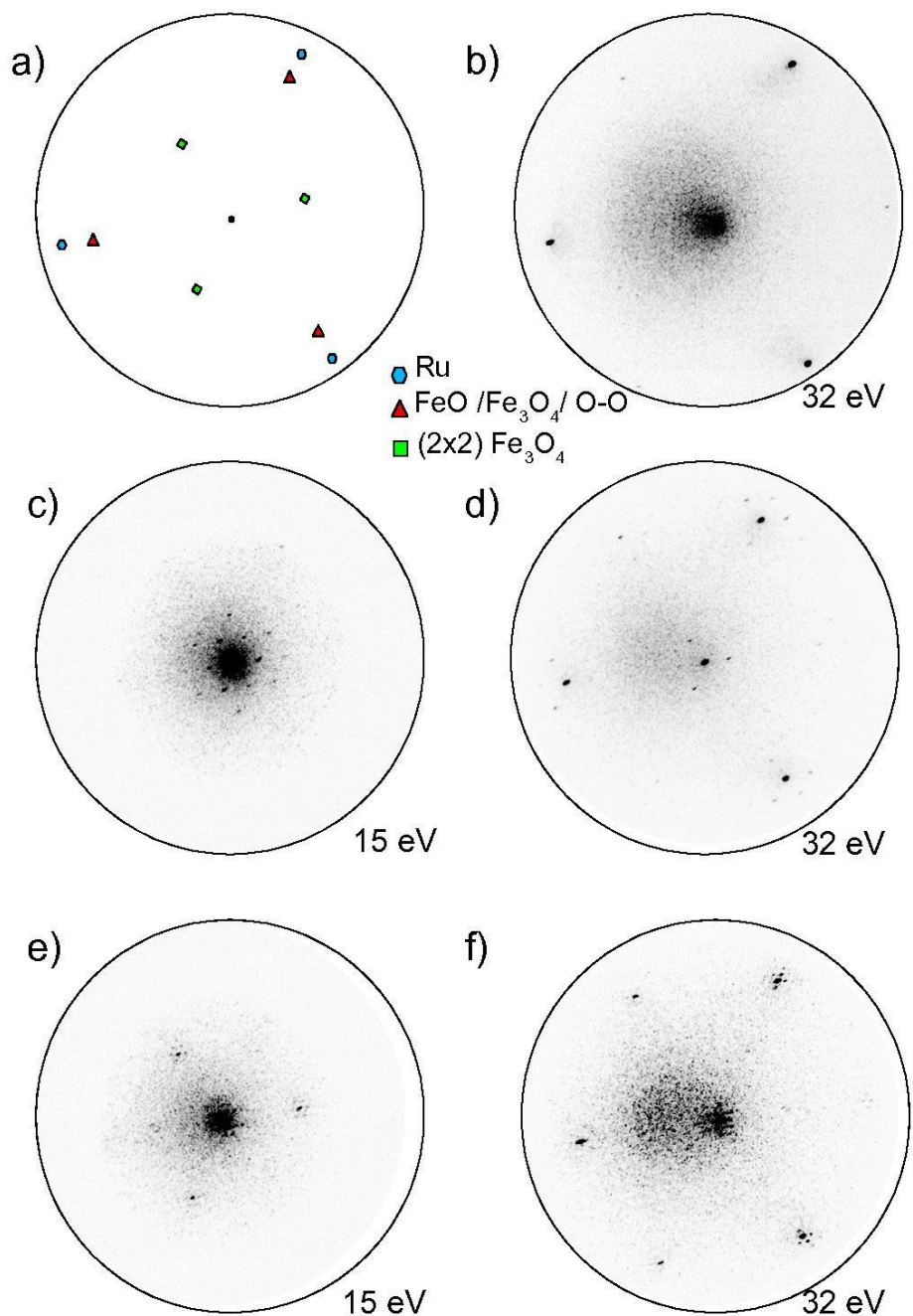

Figure 6: LEED patterns of the different iron oxides grown. Only two phases were detected. a) Schematic of the LEED patterns, showing the position of the Ru spots, and the main spots of the two iron phases, FeO and $Fe_3O_4$. b) Pattern of the Ru substrate before iron oxide growth. c-d) Patterns from the extended areas of the surface identified as FeO (see text), at two different energies (15 and 32 eV). e-f) Patterns from the triangular islands, at the same energies (15 and 32 eV). Note that the patterns at different energies have the same magnification, so they can be directly compared.

A different diffraction pattern is found when the electron beam is focused onto the second phase on the surface. This second phase first appears, in the shape of triangular islands, at coverages corresponding to more than 1 ML pure Fe (see figure 4 and figure 5). The most intense spots appear at the same positions as before (comparing figure 6d and figure 6f), indicating a periodicity of 3.02± 0.2 Å in real space. But the satellite spots appear much closer in reciprocal space, indicating a much larger mesh-size of the coincidence lattice with the substrate (54.5± 7 Å). Furthermore, at selected energies spots with a 2x2 periodicity (with satellites around them) can be detected (see figure 6e). As before, the spots are sharp, and several sets of satellite spots are detected. This again indicates the excellent crystalline order of these islands. As mentioned before, in most iron oxide films the in-plane periodicity of oxygen atoms is close to 3.0 Å. Bulk magnetite, specifically, has a value of 2.97 Å along the (111) planes. Thus, we interpret the most intense spots as those corresponding to oxygen nearest neighbour distances. The 2x2 periodicity can be understood by assuming that the real unit cell is $2\times (3.02\pm 0.2$ Å $)= 6.04\pm 0.4$ Å. This would be consistent with the notion that the oxide is either magnetite or maghemite, both of which present a 2x2 unit cell relative to the oxygen in-plane periodicity (see figure 1f). The (2x2) periodicity has been used previously to identify magnetite[16], and accordingly we also assign the phase of the triangular islands to be, most likely, magnetite. We caution, though, that in very thin films we might observe transitional structures based on the same fcc-oxygen packing with different iron ion arrangements. In addition, of course, we cannot ignore the possibility that Ru doping might affect the phases observed in these films.

We used dark-field imaging conditions to confirm that different regions on the surface give rise to the two different LEED patterns, i.e., closely spaced satellite spots (and 2x2 beams) for triangular islands and more widely separated satellite spots for FeO. In figure 7 we show both the bright field image (the image obtained from the specularly reflected electrons) and a dark field image obtained from electrons diffracted into one of the satellites that only appear in the magnetite phase. Consistent with our interpretation that they are magnetite, only the triangular islands appear bright in such an image.

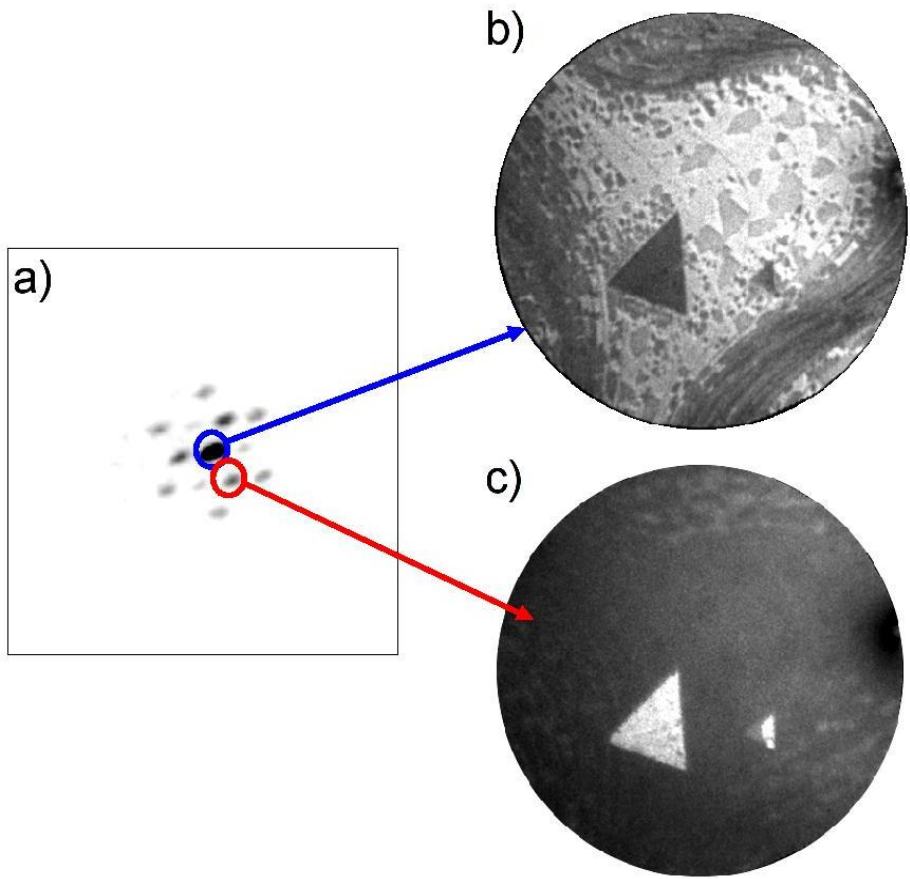

Figure 7: Comparison of dark field and bright field images of the oxide film. a) Diffraction pattern at low energy of the reciprocal space close to the specular beam (most intense beam at the center of the image). b) Bright field image of the surface. The FOV is 7µ m. c) Dark field image using a satellite spot that corresponds to the magnetite phase.

Another way to use LEEM for structural characterization is to measure electron reflectivity at very low energies. This can be done in imaging mode, by varying electron energy while recording LEEM images. In analogy to classic LEED/IV experiments, this is equivalent to measuring the intensity of the specular beam at very low energies (very low energy electron diffraction or VLEED[33,34]). Although in LEEM this is a straight-forward experiment, the interpretation of such information is not simple. Because of the poorly known behavior of the inner potential as a function of electron energy, there are uncertainties in the usual LEED multiple-scattering calculations that make them unsuitable to the 0-30 eV energy range. Nevertheless, some successful interpretations of electron reflectivity at very low energy have been achieved[29,31]. Furthermore, there are several cases were the dependence of reflectivity on electron energy has been be used to fingerprint surface structures, even in the absence of a theoretical procedure to simulate the spectra. In this "fingerprint approach", electron reflectivity has been used to locate surface structures[35] or to measure thermal adatom concentrations[36,37].

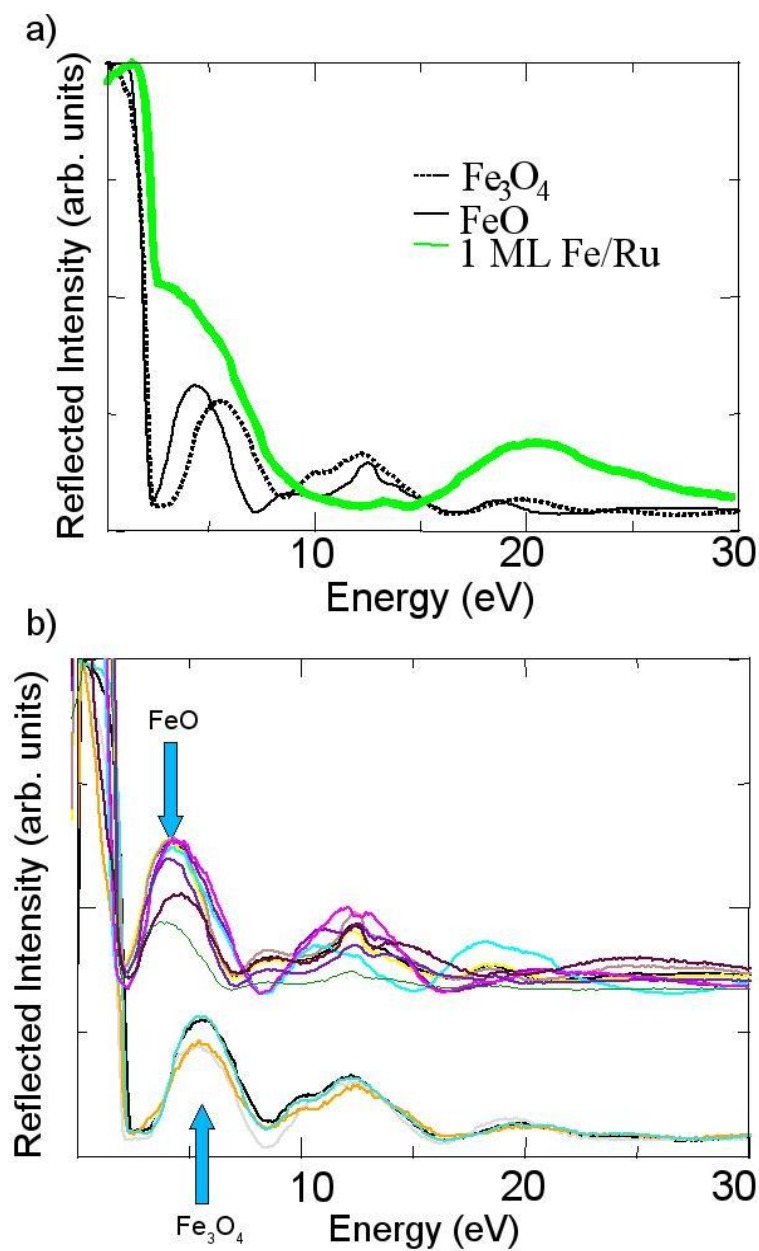

Figure 8: a) Reflectivity measurements of the oxide phases and the 1 ML Fe/Ru

surface. a) Comparison of the reflectivity on the FeO, $Fe_3O_4$, and 1ML Fe on Ru(0001). b) Reproducibility of phases grown under different experimental conditions. Curves have been offset for clarity.

We use electron reflectivity in the same way: as a fingerprint of the different film phases. Our goal is to be able to identify the surface phases through a method that does not rely on electron diffraction patterns. The motivation is to have a characterization method that can be used in our spin-polarized LEEM, which is not set up to image diffraction patterns. The low energy reflectivity of each of the iron oxide phases, and that of a monolayer of iron on ruthenium, is shown in figure 8a. The reflectivity curve of 1 ML iron on Ru is typical for metal single-crystal surfaces: after a large decrease in reflected intensity at very low energy, there is a broad peak that appears close to where single or kinematic scattering would predict the first Bragg peak, at $\approx$ 20 eV. In contrast, the iron oxide spectra in figure 8a do not show a clear peak in the $\approx$ 20 eV energy range, showing instead pronounced oscillations in the reflected electron intensity at lower energies. The phase identified as FeO has a distinct peak close to 4.3 eV, while the magnetite phase has a distinct peak at a slightly higher energy, 5.3 eV. We checked how robust these peaks are to different preparation procedures, because defect density is known to affect LEED-IV spectra of iron oxides [10]. We show in figure 8b curves acquired from the two phases grown at different temperatures, and with and without annealing. The reflectivity spectra of the magnetite phase are all very similar, independent of preparation details. In case of the FeO phase, the LEED-IV spectra show more variability as a function of different preparations. We suggest that this different behaviour could be attributed either to different film thickness, different surface terminations, or different cation distributions. However, in all FeO preparations the peak at 4.3 eV is remarkably similar, thus we can take the position of this peak into account as a useful factor to fingerprint iron oxide phases in the SPLEEM.

*3.4 Magnetic characterization by SPLEEM*

In this section we discuss magnetic measurements on iron oxide films. Wüstite is antiferromagnetic in the bulk, with a Néel temperature below room temperature (RT). The iron atoms within each (111) plane are ferromagnetically coupled, while different planes are coupled antiferromagnetically. The spin orientation is along the [111] direction, and the magnetic moment per atom is close to 4 $\mu_B$[1,38]. In contrast, magnetite is ferrimagnetic at RT: the tetrahedrally coordinated (A) and octahedrally coordinated (B) iron ions are antiferromagnetically coupled, while the interaction is ferromagnetic within each A or B lattice. The net magnetic moment per formula unit is 4 $\mu_B$. The magnetic easy axis for bulk magnetite is along [111]. For very thin films, there is an ongoing discussion on the magnetic properties of magnetite and the possible presence of magnetic dead layers. For example, recent work reports on enhanced magnetic moments of up to 7 $\mu_B$ for films thinner than 20 nm[39]. In contrast, other studies of magnetic properties of thin magnetite films grown on Pt(111) reported behaviour similar to bulk-magnetite for continuous films as thin as 7 layers thick[18].

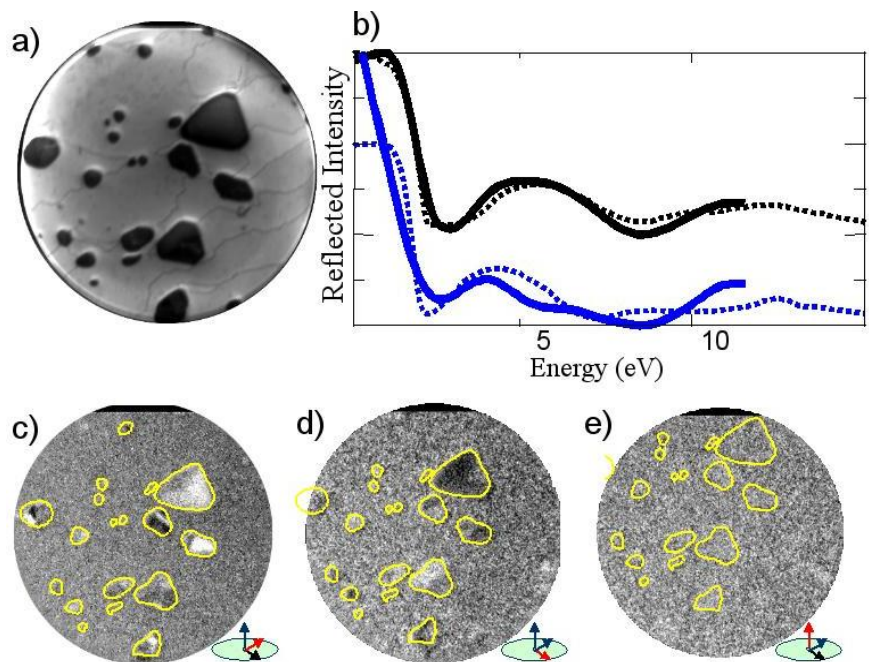

Figure 9: LEEM and SPLEEM images showing the magnetization of an oxide film prepared by annealing a 3 ML Fe film for 25 min in $10^{-6}$ torr oxygen. a) LEEM image, FOV 4$\mu$ m. b) Reflectivity curves (vertically offset for clarity) comparing films prepared in the two different LEEM systems, from the islands (black lines) and the extended areas in between (blue lines). To account for the instrumental work-function difference between both instruments, the SPLEEM-acquired reflectivity curves (continuous lines) have been rigidly shifted in energy so that the sharp drop in reflectivity falls in the same energy range as in the LEEM curves (dashed lines). c)-e) Components of the film magnetization along three orthogonal directions. The red arrows show the magnetization direction, and the yellow outline indicates the location of the islands in the magnetic contrast images.

To study in-situ the magnetization of our films we employ spin-polarized

LEEM. For a spin-polarized low-energy electron beam, the reflectivity of the sample surface depends not only on energy, topography, and other factors, but also on the relative alignment between the beam polarization and the sample magnetization. The non-magnetic contrast can be removed by acquiring pairs of images taken with reversed spin-polarizations, followed by a pixel-by-pixel subtraction of the two images. In the resulting SPLEEM images, bright (dark) contrast indicates that magnetization has a component parallel (antiparallel) to the quantization axis of the electron beam. The magnitude of the observed contrast depends on the detailed electronic band structure above the Fermi level[41]. In the absence of theory predictions, common experimental procedure consists of empirical search for magnetic contrast at different electron energies. As a rule of thumb, Fe and Co metal films give quite strong magnetic contrast, allowing detection of magnetization in thin films with image integration times of seconds or less. Ni and rare earth metals provide less contrast, requiring integration times of up to minutes.

In the SPLEEM system, the electron source is a GaAs cathode with a CsO overlayer that is sensitive to contamination. Dosing oxygen in the SPLEEM chamber, without degrading performance of the cathode, is challenging. Thus, the iron oxide films were fabricated by first depositing a number of iron layers in the SPLEEM chamber, using 500 K substrate temperature. Then the iron films were oxidized in an auxiliary chamber by heating to 900-1200 K in an atmosphere of up to $10^{-6}$ torr oxygen. Finally, we acquired reflectivity curves to characterize the films by comparing with the reflectivity spectra obtained in the other LEEM system. An oxide film obtained from a 3 ML Fe film is show in figure 9. In figure 9b the reflectivity curves acquired in the SPLEEM and the LEEM systems are compared. The reflectivity curves confirm that both the extended areas and the triangular islands correspond to the same phases identified previously by LEED, i.e., wüstite and magnetite. SPLEEM images reproduced in figure 10c-e show that there is magnetic contrast in the magnetite islands (The contrast is relatively weak, using electron beam energy 19.2 eV and integration time 5 minutes. We have tried unsuccessfully to find energies were the magnetic contrast was stronger). The magnetization direction is within the surface plane. Some islands in figure 9 are in single-domain states, and several islands have multi-domain magnetic structures.

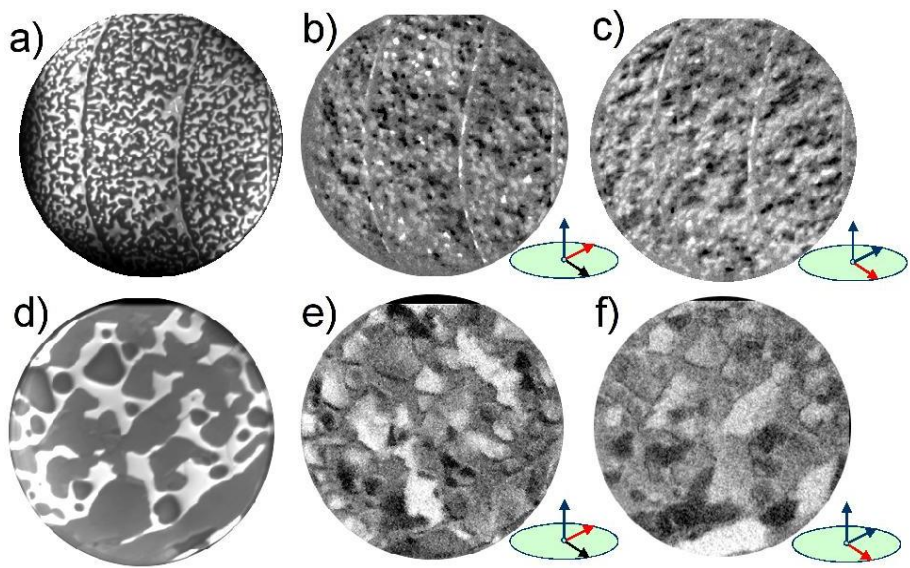

Figure 10: LEEM and SPLEEM images showing the magnetization of an iron oxide film prepared by annealing 10 ML Fe film for 15 min in $10^{-6}$ torr oxygen. a-c) LEEM and in-plane SPLEEM images, FOV 8μ m. a-c) LEEM and in-plane SPLEEM images with a smaller FOV of 4μ m, acquired after further annealing (1 hour long) in oxygen.

In thicker films (see figure 10), the magnetite islands are larger and are often interconnected. As before, no magnetic contrast is detected in the out-of-plane direction. There are still many magnetic domains within each magnetite island, with sharp domain walls between the domains. The islands thickness is not known, but given the amount of Fe deposited on the surface before oxidation and assuming they are of uniform height, we can estimate that their thickness corresponds to about 27 atomic layers. With a few nm thickness and of the order of 1000 nm lateral size, it is expected that the magnetic anisotropy would be dominated by the shape term, giving an in-plane magnetic orientation, as experimentally observed.

## 4 Summary

We have grown iron oxide films on Ru(0001) and characterized them by LEEM, using selected-area LEED, electron reflectivity, and spin-polarized LEEM. Two types of regions are clearly distinguished on the films: flat extended regions with a moiré periodicity of 18.5± 2 Å wet the substrate, while another phase with a much larger moiré periodicity of 54.5± 7 Å forms triangular islands. In both phases, there are strong diffracted peaks that correspond to 3.02± 0.2 Å. These reflections are interpreted as diffracted peaks from hexagonal lattice structures that are stabilized by the preferred bonding geometry of the oxygen anions. By comparison with previously reported STM and LEED studies[23], the flat extended regions are identified as wüstite (FeO). The second phase (triangular islands) is tentatively assigned to magnetite. The phases can be distinguished by electron reflectivity peaks that appear at 4.3 eV and 5.3 eV, in wüstite and magnetite, respectively. SPLEEM imaging shows that the magnetite islands are ferromagnetic, with in-plane domains within each island.

**Acknowledgements**

This research was partly supported by the Office of Basic Energy Sciences, Division of Materials Sciences, U.S. Department of Energy under Contracts No. DE-AC04-94AL85000 and No. DE-AC02-05CH11231, by the Spanish Ministry of Education and Science through Projects MAT2006-13149-C02-02. B.S. acknowledges support to the Spanish Ministry of Education and Science from an FPI fellowship.